\documentclass[conference]{IEEEtran}

\usepackage{amsmath}
\usepackage{indentfirst}
\usepackage{graphicx}
\usepackage{courier}
\usepackage{float}
\usepackage{flexisym} % for \textprime
\usepackage{tabularx}
\usepackage{pdfcomment}
\usepackage{algpseudocode}
\usepackage{algorithm}
\usepackage{url}
\usepackage[multiple]{footmisc}
\usepackage{caption}
\usepackage{subcaption}

\DeclareMathOperator*{\argmin}{arg\,min}
\DeclareMathOperator*{\argmax}{arg\,max}

\algnewcommand{\algorithmicgoto}{\textbf{go to}}%
\algnewcommand{\Goto}[1]{\algorithmicgoto~\ref{#1}}%

\begin{document}

\title{Task Scheduling on the Cloud with Hard Constraints}

\author{\IEEEauthorblockN{Long Thai, Blesson Varghese and Adam Barker}
\IEEEauthorblockA{School of Computer Science, University of St Andrews, Fife, UK\\
Email: \{ltt2, varghese, adam.barker \}@st-andrews.ac.uk}
}

\maketitle

\begin{abstract}
%\boldmath
Scheduling Bag-of-Tasks (BoT) applications on the cloud can be more challenging than grid and cluster environments. This is because a user may have a budgetary constraint or a deadline for executing the BoT application in order to keep the overall execution costs low. The research in this paper is motivated to investigate task scheduling on the cloud, given two hard constraints based on a user-defined budget and a deadline. A heuristic algorithm is proposed and implemented to satisfy the hard constraints for executing the BoT application in a cost effective manner. The proposed algorithm is evaluated using four scenarios that are based on the trade-off between performance and the cost of using different cloud resource types. The experimental evaluation confirms the feasibility of the algorithm in satisfying the constraints. The key observation is that multiple resource types can be a better alternative to using a single type of resource. 
\end{abstract}

\IEEEpeerreviewmaketitle

\section{Introduction}
\label{introduction}
Bag-of-Tasks (BoT) is the term given to a collection of independent and identical tasks, which can be executed in any order. BoT applications are a common way of breaking up a complex problem into smaller independent tasks in both scientific and industrial communities. BoT applications are normally executed on distributed environments in order to achieve high degrees of parallelism. BIONIC \cite{BOINC} is an example of a BoT framework, it assigns tasks to volunteer computing resources and is used in more than 80 different projects ranging from astronomy to the physical sciences\footnote{\url{https://boinc.berkeley.edu/projects.php}}. 

BoT applications are commonly executed on Grid and Cluster systems. Both environments consist of multiple interconnected machines, which are already running and shared between different organisations (Grid environment), or groups in the same organisation (cluster environment). 

Cloud computing \cite{hotcloud} is considered as a more accessible alternative as it offers resources, which a user can acquire on-demand through a pay-per-use model. However, in contrast to other forms of distributed computing environments, a cloud user has to decide which resources (instance types etc.) and how many resources need to be acquired \textbf{before} actually using and paying for them. A user cannot therefore greedily acquire as many resources as possible before deciding which ones are suitable for her application. Moreover, as cloud resources are pay-per-use the cost of the execution has to be taken into account. As a result, there is a trade-off between performance and cost: in order to have better performance, i.e. lower execution time, the cost has to be increased. Nevertheless, it is challenging to balance the trade-off so a that user can achieve the best performance with the lowest cost.

This paper explores executing BoT applications on the cloud with user defined hard constraints. Hard constraints are defined as conditions that always need to be satisfied. For example, consider the following two hard constraints: one user might want to keep the cost of execution within a certain budget constraint, while another user would prefer that the application execution must be finished within a given time frame or deadline constraint. 

In this paper, we aim to optimise the execution of BoT applications on the cloud. We investigate two scenarios in which a user provides a hard constraint in the form of a budget (the maximum amount of money that a user can spend) or a deadline (the maximum amount of time that an execution can take). If the budget constraint is given, our approach not only satisfies the constraint, but also minimises the execution time. Similarly, if the deadline constraint is provided, our approach aims to also minimise the total cost for executing the BoT. 

The contributions of this paper are as follows: i) the mathematical model for scheduling tasks on the cloud with a given hard constraint, ii) the heuristic algorithm for cost effective scheduling, and iii) the evaluation considering different trade-off between multiple options provided by Cloud providers.

The remainder of this paper is organised as follows. Section \ref{model} presents a mathematical model of the platform and the problem. Section \ref{algorithm} proposes a heuristic algorithm considering the hard constraints for executing tasks on the cloud. Section \ref{assignment} describes the scheduling of tasks. Section \ref{evaluation} presents an evaluation of the heuristic algorithm on four possible scenarios. Section \ref{relatedwork} highlights the work related to the research reported in this paper. Section \ref{conclusion} concludes this paper.

\section{Mathematical Models}
\label{model}
In this section, mathematical models that represent the cloud platform and the problem of executing the BoT on the platform given the hard constraints are considered. 

\subsection{Platform Model}
Let $IT = \{ it_1 ... it_M \}$ denote the list of $M$ types of cloud instance (for example, public cloud providers such as Amazon provide a variety of instance types\footnote{\url{http://aws.amazon.com/ec2/instance-types/}}). Each instance type $it \in IT$ can be characterised by two properties: (i) cost per hour $c_{it}$, which is the amount spent for using a Virtual Machine (VM) of an instance type in an hour, and (ii) performance $p_{it}$, which is the time taken to execute a task in seconds. Assume $T = \{ t_1 ... t_N \}$ is the list of $N$ tasks.

As the goal is to create VMs of different instance types and assign tasks to each VM, let $VM = \{ vm_1 ... \}$ be the execution plan containing the collection of VMs. For a given $vm \in VM$, the instance type is $it_{vm} \in IT$ and the tasks assigned on it are $T_{vm} \subseteq T$. A VM can be represented as $vm = (it_{vm}, T_{vm})$, which is a pair of the instance type and the tasks assigned to it. It should be noted that the upper bound of $VM$, the instance type, and the collection of tasks assigned to each VM are unknown and needs to be determined.

Normally, some amount of time is required to boot a VM into a usable state. If $st$ be this time for each VM regardless of its instance type, then the execution time of a VM is:
\begin{equation}
	exec_{vm} = \sum_{t \in T_{vm}}{p_{vm}} + st = |T_{vm}| \times p_{vm} + st
\end{equation}

Cloud VMs, for example, from public providers, such as Amazon, are charged by the hour (3600 seconds). A user pays for an hour even if only a few seconds of the hour are utilised. The cost of running a VM is:
\begin{equation}
	cost_{vm} = \lceil \frac{exec_{vm}}{3600} \rceil \times c_{it_{vm}}
\end{equation}

Since all VMs are running simultaneously, the overall execution time is the execution time of the slowest VM:
\begin{equation} \label{eq:exec}
	exec = \max_{vm \in VM}{exec_{vm}}
\end{equation}

The total cost of executing all tasks is the sum of costs of each VM represented as:
\begin{equation}
	cost = \sum_{vm \in VM}{cost_{vm}}
\end{equation}

\subsection{Problem Model}
In this paper, two hard constraints are considered. The first referred to as a \textit{`budget constraint'} is the maximum amount of money that a user is willing to spend for executing a BoT on the cloud. The second is a \textit{`time constraint'}, which is the maximum time that can be allowed for completing the execution of the BoT.

The problem of executing a BoT on the cloud given a budget constraint $B$ is to minimise the overall execution time while keeping the cost less than or equal to the budget at the same time. This is represented as follows:
\begin{equation} \label{eq:t:budget}
	\begin{split}
		&\text{minimise}		\quad	exec \\
		&\text{such that}		\quad	cost \leq B
	\end{split}
\end{equation}

Similarly, the problem of executing a BoT on the cloud given a time constraint $D$ is to minimise the total cost while keeping the overall execution time less than or equal to the deadline. This is represented as follows:
\begin{equation} \label{eq:t:deadline}
	\begin{split}
		&\text{minimise}		\quad	cost \\
		&\text{such that}		\quad	exec \leq D
	\end{split}
\end{equation}

%\subsection{Problem Modelling Using Throughput}
\subsection{Accounting for Throughput}
In order to find an optimal execution plan based on the given hard constraints, the number of VMs for each instance type and the assignment for each tasks must be decided. This is a hard problem since the number of tasks in an application are large. In this section, we simplify the problem by modelling it using \textbf{throughput}, which is the number of tasks that can be executed during a fixed period of time on a VM.

For any instance type $it \in IT$, as considered above, $p_{it}$ is the time in seconds required to execute a task. The number of tasks executed per second is $\frac{1}{p_{it}}$, and the throughput of a VM in one hour is:
\begin{equation}
	th^{3600}_{it} = \lfloor \frac{3600 - st}{p_{it}} \rfloor
\end{equation}
The floor function is applied since a task has to be fully executed by a VM.

%As $VM_{it}$ be the all VMs of type $it$, then the total throughput of instance type $it$ in one hour is:
The total throughput of all VMs of an instance type $it$ in one hour is:
\begin{equation} \label{eq:th:budget:per_it}
	TH_{it} = |VM_{it}| \times th^{3600}_{it}
\end{equation}
and the total throughput of all VMs is:
\begin{equation}
	TH = \sum_{it \in IT}{TH_{it}}
\end{equation}
%Because total throughput $TH$ represents the number of tasks executed in an hour, the performance is maximised when $TH$ is maximised.
Performance on the cloud is maximised when $TH$ is maximised.

The cost of running one VM for two hours is the same as the cost of running two VMs for one hour. Hence, it is assumed that all VMs run for no more than one hour, and the total cost is calculated as:
\begin{equation}
	cost = \sum_{it \in IT}{|VM_{it}| \times c_{it}}
\end{equation}

Given the budget constraint $B$, the problem of executing a BoT on the cloud is modelled as:
\begin{equation} \label{eq:th:budget}
	\begin{split}
		&\text{maximise}		\quad	TH \\
		&\text{such that}		\quad	cost \leq B
	\end{split}
\end{equation}

%In order to formally model the problem of executing BoT on the Cloud with deadline constraint using throughput, we first need to calculate how many tasks that a VM of an instance type $it \in IT$ can be executing \textbf{within the deadline time frame}:

If the time constraint $D$ is given, the number of tasks that can be executed by an instance type $it$ within the deadline is:
\begin{equation} \label{eq:th:deadline:per_it}
	th^D_{it} = \lfloor \frac{D - st}{p_{it}} \rfloor
\end{equation}

Given the time constraint $D$, the total throughput is:
\begin{equation}
	TH^D = \sum_{}{|VM_{it}| \times th^D_{it}}
\end{equation}
Then, taking throughput into account, the problem model is:
\begin{equation} \label{eq:th:deadline}
	\begin{split}
		&\text{minimise}		\quad	cost \\
		&\text{such that}		\quad	TH^D \geq N
	\end{split}
\end{equation}

In comparison to solving Equation \ref{eq:t:budget} (or Equation \ref{eq:t:deadline}), Equation \ref{eq:th:budget} (or Equation \ref{eq:th:deadline}) is less complex and can be solved easily since it depends on the instance types rather than the tasks.

\section{Algorithms}
\label{algorithm}
In this section, we propose a heuristic algorithm to find an execution plan based on budget and time constraints. As a starting point, a baseline solution is considered that provides a list of VMs of a single instance type and satisfies the constraints. Then, the solution is optimised to either (i) increase the performance when the budget constraint is given, or (ii) reduce the cost when the budget constraint is provided. The proposed algorithm accepts a single constraint at a time and simultaneously considering multiple constraints will be investigated in the future.

\subsection{Select the Most Cost Effective Instance Type}
%\subsubsection{Budget Constraint} 
For an instance type $it \in IT$, the number of VMs affordable under a budget constraint $B$ is:
\begin{equation} \label{eq:budget:num}
	|VM_{it}|^B = \lfloor \frac{B}{c_{it}} \rfloor
\end{equation}

As presented by Equation \ref{eq:th:budget:per_it}, $TH_{it}$ is the throughput per hour of \textbf{one} VM of $it$. Hence, the total throughput produced by $it$, based on cost constraint $B$ is:
\begin{equation}
	TH^B_{it} = |VM_{it}|^B \times TH_{it}
\end{equation}

Hence, for a given budget, the most cost effective instance type $it^B$ is:
\begin{equation} \label{eq:budget:it}
	it^B = \argmax_{it \in IT}{TH^B_{it}}
\end{equation}
that results in the highest total throughput.

%\subsubsection{Deadline Constraint} 
%With the deadline constraint, the number of tasks which can be executed by an instance type $it$ is $th^D_{it}$, as shown by \ref{eq:th:deadline:per_it}. Hence, the number of VMs required is:

The number of tasks that can be executed on an instance type under a time constraint is $th^D_{it}$ (refer Equation \ref{eq:th:deadline:per_it}). So, the number of VMs required to satisfy the time constraint is:
\begin{equation} \label{eq:deadline:num}
	|VM_{it}|^D = \lceil \frac{N}{th^D_{it}} \rceil
\end{equation}

The cost for the VMs is:
\begin{equation}
	cost^D_{it} = |VM_{it}|^D \times c_{it}
\end{equation}

The most cost effective instance type is:
\begin{equation} \label{eq:deadline:it}
	it^D = \argmin_{it \in IT}{cost^D_{it}}
\end{equation}

\subsection{Optimise Instance Type Selection Algorithm}

%Equations \ref{eq:budget:it} and \ref{eq:deadline:it} aim to find the most cost effective instance type given the budget or cost constraints. However, by applying them, it is possible to build a execution plan with \textbf{only one instance type}. On the other hand, a plan which multiple types can performance better with cheaper cost.

The execution plan by applying budget or cost constraints is generated using a single instance type. However, when multiple instance types are utilised in the execution plan, it is possible to obtain better performance and also reduce costs further.

Algorithm \ref{al} optimises the execution plan by replacing VMs initially present in the execution plan with VMs of other instance types. This not only increases the throughput but also reduces the costs while satisfying the constraints.

\begin{algorithm}
	\caption{Optimise Instance Type Selection}
	\label{al}
	\begin{algorithmic}[1]
		\Function{OPTIMISE}{$IT, it_0, B, D, num, minCost$}
			\State $exec \gets D - st$
			\If{$num_{it_0} = 0 \vee num_{it_0} < num$}								\label{al:check}
				\State \Return
			\EndIf
			\State $(it', th_{m}, num_{rped}, num_{rping}) \gets (NULL, 0, 0, 0)$
			\For{$it_1 \in \{ it \in IT \mid it \neq it_0 \}$}
				\If{$c_{it_1} > B + num \times c_{it_0}$}							\label{al:check_cost}
					\State \textbf{continue}
				\EndIf
				\State $num' \gets 0$
				\If{$c_{it_1} > B$}													\label{al:check_budget}
					\State $num' \gets num$											\label{al:replaced_num}
				\EndIf
				\State $num\_vms \gets \lfloor \frac{B + num' \times c_{it_0}}{c_{it_1}} \rfloor$		\label{al:replacing_num}
				\If{$minCost = TRUE \wedge num\_vms \times c_{it_1} = B + num' \times c_{it_0}$}	\label{al:check_minimise_cost}
					\State $num\_vms \gets num\_vms - 1$
				\EndIf
				\State $th' \gets th - \lfloor \frac{exec}{th_{it_0}} \times num' \rfloor + \lfloor \frac{exec}{th_{it_1}} \times num\_vms \rfloor$	\label{al:new_throughput}

				\If{$minCost = TRUE \wedge th' < |T|$}								\label{al:check_deadline_throughput}
					\State \textbf{continue}
				\ElsIf{$th' < th$}														\label{al:check_budget_throughput}
					\State \textbf{continue}
				\ElsIf{$th' > th_{m} \vee (th' = th_{m} \wedge c_{it_1} < c_{it'})$}	\label{al:select_replacing}
					\State $(it', th_{m}, num_{rped}, num_{rping}) \gets (it_1, th', num', num\_vms)$
				\EndIf
			\EndFor
			\If{$it' = NULL$}
				\State $REPLACE(IT, it_0, B, D, num + 1, minCost)$	\label{al:continue_fail}
			\Else
				\State $num_{it_0} \gets num_{it_0} - num_{rped}$		\label{al:update_replaced}
				\State $num_{it'} \gets num_{it'} + num_{rping}$		\label{al:update_replacing}
				\If{$minCost = TRUE$}															\label{al:update_budget_zero}
					\State $B \gets 0$
				\Else
					\State $B \gets B + num_{rped} \times c_{it_0} - num_{rping} \ times c_{it'}$	\label{al:update_budget}
				\EndIf
				\State $REPLACE(IT, it_0, B, D, num, minCost)$			\label{al:continue_succeed}
			\EndIf
		\EndFunction
	\end{algorithmic}
\end{algorithm}

The inputs to Algorithm \ref{al} are the list of instance types, the selected instance type $it_0$ (which is obtained from Equation \ref{eq:budget:it} or Equation \ref{eq:deadline:it}), the \text{remaining budget} (which will be explained shortly), the time constraint, the number of VMs to be replaced and a boolean flag indicating if the goal is to minimise cost ($TRUE$ for minimising cost and $FALSE$ maximising throughput). 

The boolean flag determines the calculation of the remainder of the  budget. If the goal is to maximise throughput, then the budget is the difference between the cost constraint initially provided and the current cost of execution. On the other hand, if the goal is to minimise cost, then no more VMs can be added since the budget is depleted. 

%It should be noted that the remaining budget is calculated differently depending on difference scenario: if the goal is to maximise throughput, the remaining budget is the difference between the given budget constraint and the cost of execution. On the other hand, if the goal is to minimise cost, its value is always $0$, i.e. there is no budget to add more VMs.

Algorithm \ref{al} is recursive and in each iteration, either the performance is improved or the cost is reduced without violating the given constraint. The algorithm terminates when the number of VMs of $it_0$ is either zero or less than the number of VMs to be replaced (Line \ref{al:check}).

The algorithm firstly loops through all instance types except $it_0$. Instance types that cannot be afforded within the budget are ignored (Line \ref{al:check_cost}). The allowance to add new VM(s) is the sum of the remaining budget and the cost of VMs of $it_0$ which are allowed to be replaced (if the remaining budget is not enough, some VMs of $it_0$ will have be removed to not exhaust the budget).

Then the number of VMs to be replaced is calculated (Lines \ref{al:check_budget} and \ref{al:replaced_num}). After that, based on the allowance, the number of replacing VM is calculated (Line \ref{al:replacing_num}). If the goal is to minimise cost and the cost of additional VMs is \textbf{exactly equal} to the allowance, the number of replacing VM has to be decreased by one so that its cost can be lower than the allowance (Line \ref{al:check_minimise_cost}).

Next, the resulting throughput is calculated by adding the additional throughput from the newly added VMs and the current throughput. VMs of $it_0$ are removed, and their throughput is deducted (Line \ref{al:new_throughput}).

If the goal is to minimise the total cost, the new throughput must not be less than the total number of tasks as all tasks must be executed within the deadline (Line \ref{al:check_deadline_throughput}). On the other hand, if the goal is to maximise throughput, the new throughput cannot be less than the current throughput (Line \ref{al:check_budget_throughput}).

All replacement instance types are compared; the instance type with the highest throughput and lowest cost is selected (Line \ref{al:select_replacing}). Then, VMs of the new instance type are added and VMs of $it_0$ are removed if required. This process is performed by changing the number of VMs of each instance type (Lines \ref{al:update_replaced} and \ref{al:update_replacing}). The remaining budget is updated (Line \ref{al:update_budget}); the remaining budget is always zero (Line \ref{al:update_budget_zero}) if the goal is to minimise cost.

The algorithm continues execution with updated values (Line \ref{al:continue_succeed}). However, if no replacing instance types are found, the number of VMs of type $it_0$ is increased by one (Line \ref{al:continue_fail}). Each iteration can result in the following: (i) increase the overall throughput, or (ii) decrease the total cost, or (iii) increase the number of VMs of $it_0$ to be replaced by one.

\section{Assign Tasks to VMs}
\label{assignment}
The result of Algorithm \ref{al} is a list of VMs of different instance types. However, the VMs do not have any tasks assigned to them. Therefore, we propose an additional algorithm (refer to Algorithm \ref{al:assign}) for assigning tasks to VMs. The algorithm finds a VM in the list which can complete the execution of the task in the lowest time if a task were assigned to it. This ensures that the overall execution is as low as possible.

For example, given two instance types whose performances are $5$ and $8$, and assuming there are two VMs of the first type and one VM of the second type, the execution time for each VM is $10$, $12$ and $9$. If a task is added to each VM, their new execution will be $10 + 5 = 15$, $12 + 5 = 17$ and $9 + 8 = 16$. Hence, a new task should be added to the first VM, whose performance is $5$ and current execution time is $10$, so that the overall execution time after the assignment is lower than the other options.

\begin{algorithm}
	\caption{Assign Tasks to VMs}
	\label{al:assign}
	\begin{algorithmic}[1]
		\Function{ASSIGN}{$T, VM$}
			\For{$t	\in T$}
				\State $vm_0 \gets NULL$
				\State $exec \gets 0$
				\For{$vm \in VM$}
					\State $exec' \gets exec_{vm} + p_{it_{vm}}$
					\If{$vm_0 = NULL \vee exec' < exec$}
						\State $vm_0 \gets vm$
						\State $exec \gets exec'$
					\EndIf
				\EndFor
				\State $T_{vm_0} \gets T_{vm_0} \cap \{ t \}$
			\EndFor
			\State \Return $VM$
		\EndFunction
	\end{algorithmic}
\end{algorithm}

% The assignment performed by Algorithm \ref{al:assign} has the complexity $N \times |VM|$ since it considers all VMs for each task. Although the complexity is high (dependent on the total number of tasks), the algorithm can be implemented and executed in reasonable time.

\section{Experimental Evaluation}
\label{evaluation}
Our approach of scheduling tasks on the cloud based on a given budget or cost constraint is evaluated in this section. The evaluation considers four different scenarios which are based on the difference in cost and performance between different instance types.

\subsection{Performance Gain vs Cost Increase}
One important criteria for selecting a cloud instance is the trade-off between performance and cost. For example, how much quicker does a task execute when it is moved from one instance type to another with a higher cost. Given an instance type $it_0$, the trade-off when employing a more expensive instance type $it_1$, where $c_{it_1} > c_{it_0}$ can be calculated as: $to_{it_1, it_2} = \frac{p_{it_0}}{p_{it_1}} / \frac{c_{it_1}}{c_{it_0}}$ (ratio of the change in performance and change in cost). 

There are three cases for the trade-off between performance and cost:
\begin{itemize}
	\item Fair trade-off ($to = 1$), when the performance gain is equal to the increase in cost. When the trade-off is fair, it does not make much difference between using expensive or cheaper instances.
	\item Cost-effective trade-off ($to > 1$), when the performance gain is more than the monetary increase. It is profitable to use an expensive instance type.
	\item Cost-ineffective trade-off ($to < 1$), when the performance gain is less than the monetary increase. In this case, it is advisable to use cheap instance types.
\end{itemize}

The trade-off is highly specific to a user's application and instance type. For example, if an application can use only one CPU core, using expensive instance types with multiple cores may not be cost-effective. On the other hand, instance types with more cores, each of which has higher clock speed, can be beneficial to a parallel application.

In order to effectively select the optimal execution plan for an application, the trade-off between performance gain and monetary increase must be taken into account. It is more beneficial to have many VMs of the cheap instance type if the trade-off is cost-ineffective. On the other hand, with the cost-effective trade-off, a user can opt for VMs of expensive instance types.

\subsection{Setup}
This section presents a comparison between using only one instance type or combining multiple instance types. We compare two approaches, the first one is simple and only uses the most cost-effective instance type selected by Equations \ref{eq:budget:it} or \ref{eq:deadline:it} while the other applied Algorithm \ref{al} to use a combination of multiple instance types.

% In order to evaluate Algorithm \ref{al} in finding execution plan for BoT application on the cloud based on given budget or deadline constraints
Four different scenarios corresponding to the three trade-off cases and an additional mixed trade-off (cost-effective and cost-ineffective VM types were used) case were considered.

For each scenario, we used 10 different values of budget and deadline constraints and an application with 1000 tasks. The instance start up time, i.e. $st$, is set to $10$ seconds.

\subsubsection{Scenario 1 - Fair Trade-off} Table \ref{tab:s1} shows that the performance and cost of an instance increases in the same ratio. For example, $it_2$ is two times more expensive than $it_1$ and the time it takes to execute an application is half of $it_1$'s.

\begin{table}
	\begin{tabularx}{0.45\textwidth}{ | X | X | X | }
		\hline
		Instance Type	&	Cost	&	Performance	\\
		\hline
		$it_1$			&	1		&	32			\\
		\hline
		$it_2$			&	2		&	16			\\
		\hline
		$it_3$			&	4		&	8			\\
		\hline
		$it_4$			&	8		&	4			\\
		\hline
		$it_5$			&	16		&	2			\\
		\hline
	\end{tabularx}
\caption{Fair Trade-off}
\label{tab:s1}
\end{table}

\subsubsection{Scenario 2 - Cost-ineffective Trade-off} Table \ref{tab:s3} shows that the performance gain is lower than the monetary increase. For example, $to_{it_1, it_2} = 0.9$

\begin{table}
	\begin{tabularx}{0.45\textwidth}{ | X | X | X | }
		\hline
		Instance Type	&	Cost	&	Performance	\\
		\hline
		$it_1$			&	1		&	32			\\
		\hline
		$it_2$			&	2		&	18			\\
		\hline
		$it_3$			&	4		&	10			\\
		\hline
		$it_4$			&	8		&	6			\\
		\hline
		$it_5$			&	16		&	4			\\
		\hline
	\end{tabularx}
\caption{Cost-ineffective Trade-off}
\label{tab:s3}
\end{table}

\subsubsection{Scenario 3 - Cost-effective Trade-off} Table \ref{tab:s2} shows that the performance gain is more than the monetary increase. For example, $to_{it_1, it_2} = 1.07$.

\begin{table}
	\begin{tabularx}{0.45\textwidth}{ | X | X | X | }
		\hline
		Instance Type	&	Cost	&	Performance	\\
		\hline
		$it_1$			&	1		&	32			\\
		\hline
		$it_2$			&	2		&	15			\\
		\hline
		$it_3$			&	4		&	7			\\
		\hline
		$it_4$			&	8		&	3			\\
		\hline
		$it_5$			&	16		&	1			\\
		\hline
	\end{tabularx}
\caption{Cost-effective Trade-off}
\label{tab:s2}
\end{table}

\subsubsection{Scenario 4 - Mixed Trade-off} Table \ref{tab:s4} shows instance types with cost-effective and cost-ineffective trade-off. We obtained the performances by executing the genome pattern searching application on five Amazon instance types. For example, $to_{M3.Medium, C3.Large} = 2.2$ and $to_{C3.Large, M3.Large} = 0.7$.

\begin{table}
	\begin{tabularx}{0.45\textwidth}{ | X | X | X | }
		\hline
		Instance Type	&	Cost	&	Performance	\\
		\hline
		M3.Medium			&	0.077		&	87.37			\\
		\hline
		C3.Large			&	0.12		&	25.33			\\
		\hline
		M3.Large			&	0.154		&	27.08			\\
		\hline
		C3.Xlarge			&	0.239		&	12.7			\\
		\hline
		M3.Xlarge			&	0.308		&	13.79			\\
		\hline
	\end{tabularx}
\caption{Mixed Trade-off}
\label{tab:s4}
\end{table}

\subsection{Results}

% Instead of showing the actual execution time when the budget constraint is given and actual cost when the deadline is given, we present the ratio between two approaches: using only the most cost-effective instance type and using the combination of different ones. When the budget constraint is given, the result is the ratio between execution time

The four scenarios (each scenario took ten values for budget and cost constraints) considered above were simulated on a custom built simulation framework developed using Scala. The framework took as input the cost and performance of the instances considered in Table \ref{tab:s1} - \ref{tab:s4}. The framework then executed Algorithm \ref{al} to generate an execution plan. The plan was then executed and the resulting cost and performance were found to satisfy the constraints. 

Instead of demonstrating the results of overall cost and execution time, we compared two approaches (one using single instance types and the second using multiple instance types) by taking the ratio of their results. When the budget (or deadline) constraint is given, the ratio between execution times (or actual costs) of using only the most cost-effective instance type and using the combination of different ones is noted. Both approaches have the same performance if the ratio was equal to 1 and when multiple instance types are used the performance is better if the ratio was greater than 1. Single instance types perform better when the ratio was less than 1.

\begin{figure*}
	\centering
		\begin{subfigure}[b]{0.49\textwidth}
			\includegraphics[width=\textwidth]{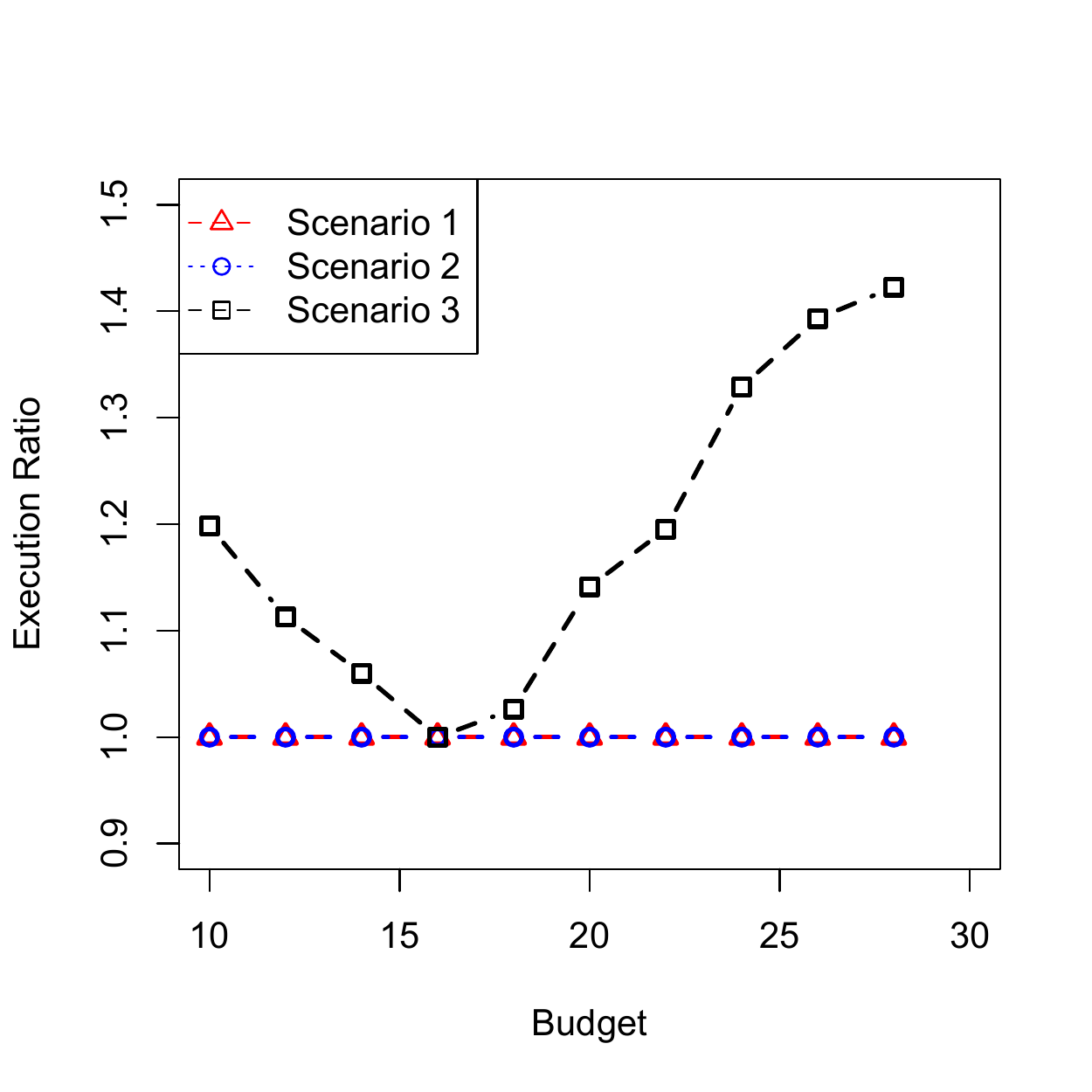}
			\caption{Results for budget constraint}
			\label{fig:result_budget_123}
		\end{subfigure}
			\begin{subfigure}[b]{0.49\textwidth}
			\includegraphics[width=\textwidth]{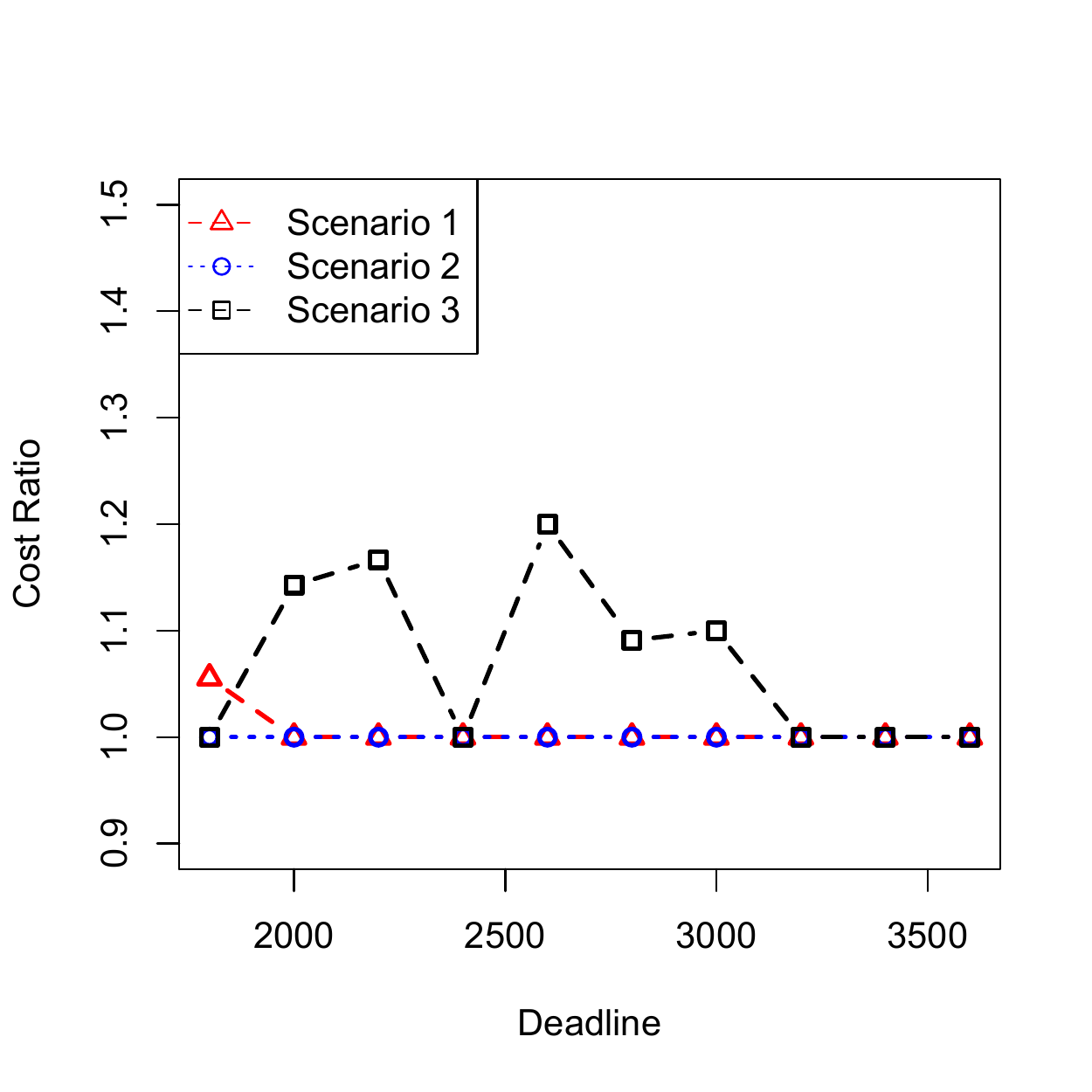}
			\caption{Results for deadline constraint}
			\label{fig:result_deadline_123}
		\end{subfigure}
		\caption{Comparison between two approaches for Scenarios 1, 2 and 3} \label{fig:result_123}
\end{figure*}

Figure \ref{fig:result_123} presents the results for Scenarios 1, 2 and 3. In scenario 1 and 2, two approaches behaved similarly most of the time. It is because they both used VMs of the cheapest instance type, which had either the same (scenario 1) or more (scenario 2) performance per cost in comparison with the remaining instance type. Hence, the remaining budget was not enough to add any more VMs of other instance types. The only time when the second approach out-performed the first was when deadline was 1800 seconds in Scenario 1 (Figure \ref{fig:result_deadline_123}). It can be explained by looking into the total number of VMs used by each approach: if only one instance type was used, an execution plan contained 19 VMs of $it_1$, whose cost was 1. However, Algorithm \ref{al} replaced 17 VMs of type $it_1$ with one VM of $it_5$, whose cost was 16. As the result, the cost could be reduced, however, the execution time increased from 1706 to 1790 which was still below the deadline constraint.

On the other hand, in Scenario 3 in which the trade-off was cost-effective, the second approach out-performed the first approach most of the time (black and squared lines in Figure \ref{fig:result_123}). It can be explained as follow: due to the cost-effectiveness, the most expensive instance type was selected by both Equations \ref{eq:budget:it} and \ref{eq:deadline:it}. While the first approach only tried to create as many VMs of the selected instance type as possible, the second approach took advantage of the remaining budget by either adding or replacing the existing VMs with VMs of other types.

\begin{figure*}
	\centering
		\begin{subfigure}[b]{0.49\textwidth}
			\includegraphics[width=\textwidth]{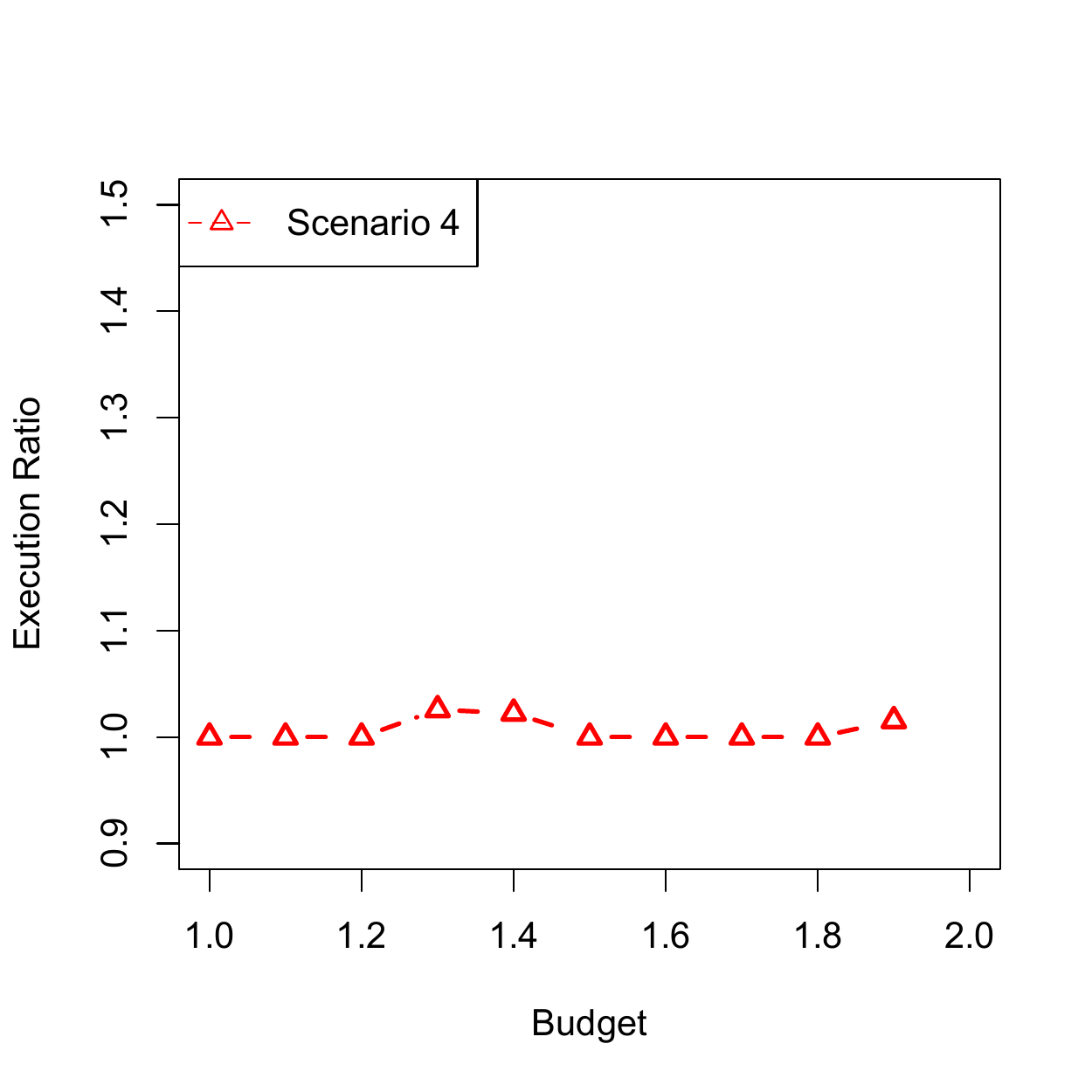}
			\caption{Results for budget constraint}
			\label{fig:result_budget_4}
		\end{subfigure}
			\begin{subfigure}[b]{0.49\textwidth}
			\includegraphics[width=\textwidth]{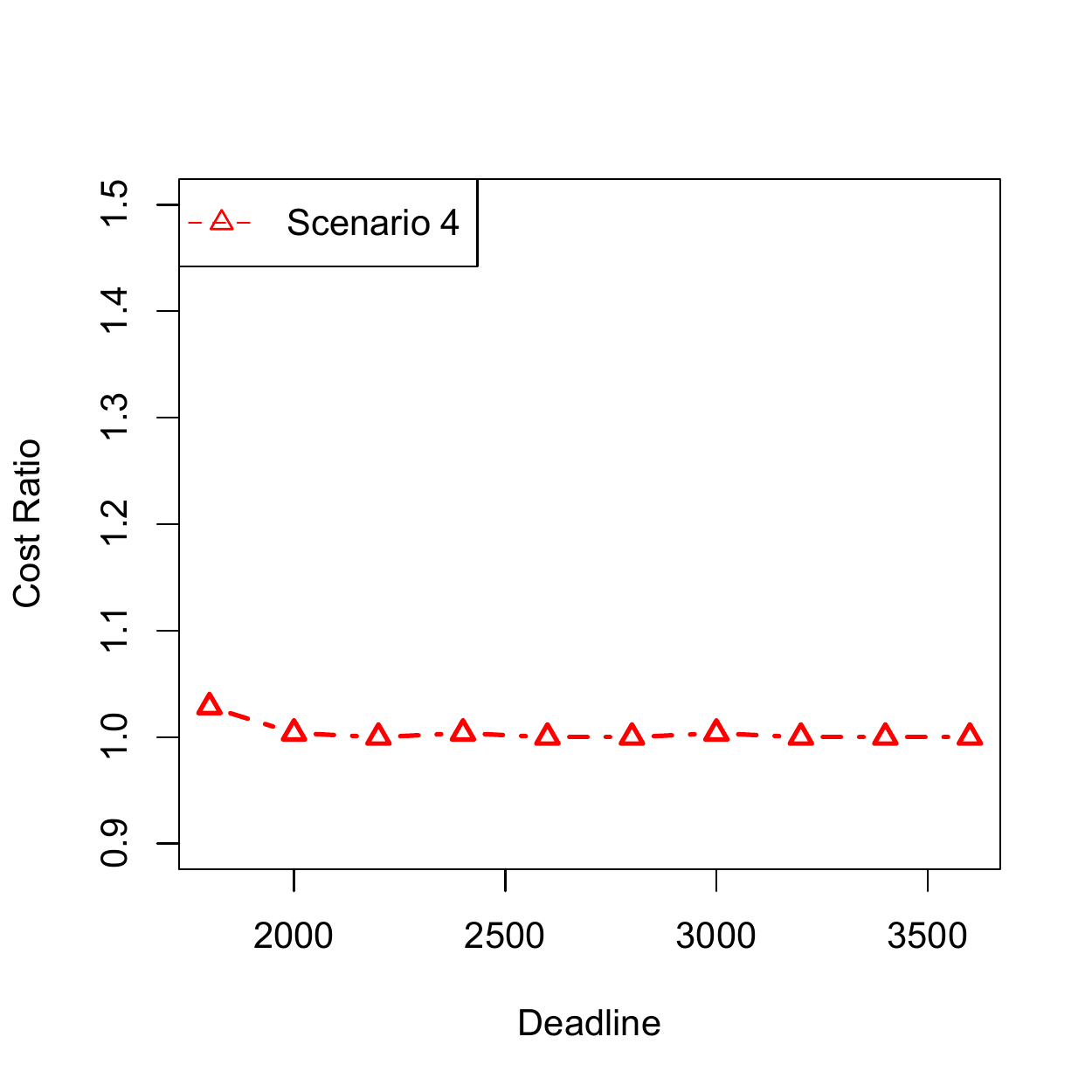}
			\caption{Results for deadline constraint}
			\label{fig:result_deadline_4}
		\end{subfigure}
		\caption{Comparison between two approaches for Scenario 4} \label{fig:result_4}
\end{figure*}

Figure \ref{fig:result_4} presents the result for the mixed trade-off scenario. It is similar to Scenario 1 or 2 in that there is no significant difference between using only one or a combination of different instance types. This can be explained as follows: two most cost-effective instance types are C3.Large and C3.Xlarge, which is understandable since they both were CPU optimised instance type and the genome search application is CPU intensive. As a result, either instance was always selected by Equations \ref{eq:budget:it} and \ref{eq:deadline:it}. Since Algorithm \ref{al} optimises the execution, it never selects instance types M3.Large or M3.Xlarge since they are more expensive and has poorer performance when compared to C3.Large and C3.Xlarge. C3.Large and C3.Xlarge demonstrate fair trade-off ($to_{C3.Large, C3.Xlarge} = 1.002$); consequentially, Scenario 4 is similar to Scenario 1. Even though VMs of M3.Medium type could be added due to the remaining budget, there was no significant improvement in performance since the instance performs poorly.

The experiments show that the hard constraints can be satisfied by the heuristic algorithm proposed in this paper. The results from the experiments show that single instance types can sometimes be as effective as multiple instance types. As a general trend, we note that when there is a cost effective trade-off multiple instance types perform better than a single type of instance. 

\section{Related Work}
\label{relatedwork}
There is extensive research focusing on executing BoT in the Grid environment. The MyGrid framework \cite{Cirne:2003:ICPP} supports the execution of BoT on the distributed environment and improves performance by replicating tasks. Similarly, algorithms to assign a collection of tasks to grid resources in order to minimise execution time has been developed \cite{Maheswaran:1999:DMS}. Scheduling tasks based on deadline constraints in the grid \cite{Takefusa:2001:HPDC} and based on the location of input for each task \cite{Ranganathan:2002:DCD} are considered. The execution of independent but file-sharing tasks is investigated and a heuristic algorithm to achieve better performance in comparison to greedy approaches is proposed \cite{Kaya:TransPDS:2006}. Scheduling tasks while satisfying both deadline and budget constraints are considered assuming that each task requires distributed data at multiple sources \cite{Venugopal:2005:ICA3PP}. Scheduling multiple BoTs is another avenue that is investigated \cite{Bertin:2008:Grid,Anglano:2008:IPDPS}. Executing parallel tasks on a single machine is presented as an alternative approach to improve performance \cite{Benoit:TransComp:2010}.

For clusters, Hadoop's YARN \cite{Vavilapalli:2013:AHY} and Mesos \cite{Hindman:2011:MPF} are resource management frameworks that allocate compute resources to applications in order to maximise performance. Task scheduling systems such as Apollo \cite{Boutin:2014:OSDI} and Omega \cite{Schwarzkopf:2013:OFS} predict the execution time of each task on a resource and use this prediction for assigning tasks. Sparrow \cite{Ousterhout:2013:SDL} performs task assignment by evaluating the length of a waiting queue of tasks to be executed on a resource. 

The cost factor is usually not taken into account while scheduling tasks on Grid and Cluster environments. On the other hand, in cloud computing, a user has to pay for the resources employed, and therefore, the monetary cost must be considered. Numerous attempts have been made to model costs related to executing BoTs on different platforms. However, these models either take into account the cost for transferring data and execution of individual tasks \cite{Venugopal:2005:ICA3PP} or are auction-based models which may not be most suited for the cloud \cite{Sulistio:2005:SBAC}.

Recently, many researchers have started to apply cloud computing for executing BoT applications. Statistical learning and constraint solvers are used to maximise the execution performance while satisfying a budget constraint \cite{Oprescu:2010:CloudCom}. Deadline constraints have also been investigated by considering the workload of VMs \cite{Mao:2010:GRID}. Methods for optimising the cost and performance on multiple clouds \cite{Farahabady:2012:PDCAT} and scheduling algorithms are considered \cite{Gutierrez-Garcia:2013:FHA}. In our previous work, we investigated the trade-off between performance and cost when executing Bag-of-Distributed-Tasks on the cloud and proposed a method to find an execution plan based on a given budget constraint \cite{Thai:2014:CloudCom}.

The research presented in this paper distinguishes itself from the current state-of-the-art in many ways. First of all, it does not put a limit on the number of cloud resources; most research assume a limit, for example, \cite{Oprescu:2010:CloudCom,Farahabady:2012:PDCAT,Gutierrez-Garcia:2013:FHA,Thai:2014:CloudCom}. The resource limit is defined by either budget or deadline constraints assuming the availability of unlimited number of resources. Moreover, our approach focuses not only on resource provisioning, but also task scheduling; this is often not considered in other research, for example, \cite{Oprescu:2010:CloudCom,Mao:2010:GRID}. Task scheduling offers the flexibility of controlling the execution of BoT on the cloud. 

\section{Conclusion}
\label{conclusion}
In comparison to other distributed environments, such as the grid and the cluster, scheduling tasks on the cloud is complex since (i) there are multiple types of resources with different performance and varying costs offered on the cloud, and (ii) a user can impose a budget or a deadline constraint. It is challenging to make a decision on the type or combination of resources that can satisfy the constraints.    

To address the above challenge, in this paper, we proposed a heuristic algorithm to schedule tasks or Bag-of-Tasks (BoT) on the cloud such that the hard constraints imposed by a user can be satisfied. The algorithm first generates an execution plan comprising the most cost effective resource and then modifies the initial plan with different resource types. We evaluated the algorithm on four scenarios which were developed by taking into account the trade-off between performance and cost of the cloud resources. If the trade-off was either fair or cost-ineffective, there was not much difference between using a single or multiple types of resources. However, if the trade-off was cost-effective, a combination of different resources was able to reduce the cost and/or increase the performance. The experiments confirms that the budget and deadline constraints can be satisfied.  

In the future, we plan to generalise our approach so that it can be applied for multiple applications. Moreover, dynamic scheduling and resource provisioning will be investigated so that the failure of a virtual machine can be handled.

\section*{Acknowledgment}

This research is supported by the EPSRC grant `Working Together: Constraint Programming and Cloud Computing' (EP/K015745/1), a Royal Society Industry Fellowship `Bringing Science to the Cloud', an Impact Acceleration Account (IAA) grant and an Amazon Web Services (AWS) Education Research Grant.

\bibliographystyle{ieeetr}
\bibliography{references}

% that's all folks
\end{document}